\documentclass[aps,prd,twocolumn,showpacs,superscriptaddress]{revtex4}

\newcommand{\ba}{\begin{array}}
\newcommand{\ea}{\end{array}}
\newcommand{\beqa}{\begin{eqnarray}}
\newcommand{\eeqa}{\end{eqnarray}}

\begin{document}

\title{Supersymmetric One-family Model without Higgsinos}
\author{Jes\'us M. Mira} 
\affiliation{Instituto de F\'\i sica, Universidad de Antioquia,
A.A. 1226, Medell\'\i n, Colombia}
\author{William A. Ponce}
\affiliation{Instituto de  F\'\i sica, Universidad de Antioquia,
A.A. 1226, Medell\'\i n, Colombia}
\author{Diego A. Restrepo}
\affiliation{Instituto de  F\'\i sica, Universidad de Antioquia,
A.A. 1226, Medell\'\i n, Colombia}
\author{Luis A. S\'anchez}
\affiliation{Instituto de  F\'\i sica, Universidad de Antioquia,
A.A. 1226, Medell\'\i n, Colombia}
\affiliation{Escuela de F\'\i sica, Universidad Nacional de Colombia,
A.A. 3840, Medell\'\i n, Colombia}

\begin{abstract}
The Higgs potential and the mass spectrum of the $N=1$ supersymmetric
extension of a recently proposed one family model based on the local gauge
group $SU(3)_C\otimes SU(3)_L\otimes U(1)_X$, which is a subgroup of the
electroweak-strong unification group $E_6$, is analyzed. In this model the
slepton multiplets play the role of the Higgs scalars and no Higgsinos are
needed, with the consequence that the
sneutrino, the selectron and six other sleptons play the role of the
Goldstone bosons. We show how the $\mu$ problem is successfully addressed  
in the context of this model which also predicts the existence of a light 
CP-odd scalar.  
\end{abstract}
\pacs{12.60.Jv, 12.60.Fr, 12.60.Cn}

\maketitle

\section{\label{sec:intr}Introduction}
In spite of the remarkable experimental success of our leading theory of
fundamental interactions, the so-called Standard Model (SM) based on the
local gauge group $SU(3)_c\otimes SU(2)_L\otimes U(1)_Y$ \cite{sm}, it
fails to explain several issues like hierarchical fermion masses and
mixing angles, charge quantization, CP violation, replication of families,
among others. These well known theoretical puzzles of the SM have led to
the strong belief that the model is still incomplete and that it must be
regarded as a low-energy effective field theory originating from a more
fundamental one.  Among the unsolved questions of the SM, the elucidation
of the nature of the electroweak symmetry breaking remains one of the most
challenging issues.  If the electroweak symmetry is spontaneously broken
by Higgs scalars, the determination of the value of the Higgs mass $M_H$
becomes a key ingredient of the model. By direct search, LEP-II has set an
experimental lower bound of $M_H\geq 114.4$ GeV \cite{lep2}.

Close after the proposal of the SM many scenarios for a more fundamental
theory have been advocated in several attempts for answering the various
open questions of the model. All those scenarios introduce theoretically
well motivated ideas associated to physics beyond the SM \cite{ext}.  
Supersymmetry (SUSY) is considered as a leading candidate for new physics.
Even though SUSY does not solve many of the open questions, it has several
attractive features, the most important one being that it protects the
electroweak scale from destabilizing divergences, that is, SUSY provides
an answer to why the scalars remain massless down to the electroweak scale
when there is no symmetry protecting them (the ``hierarchy problem"). This
has motivated the construction of the Minimal Supersymmetric Standard
Model (MSSM) \cite{mssm}, the supersymmetric extension of the SM, that is
defined by the minimal field content and minimal superpotential necessary
to account for the known Yukawa mass terms of the SM. At present, however,
there is no experimental evidence for Nature to be supersymmetric.

In the MSSM it is not enough to add the Higgsino to construct the left
chiral Higgs supermultiplet. Because of the holomorphicity of the
superpotential and the requirement of anomaly cancellation, a second Higgs
doublet together with its superpartner must be introduced. The two Higgs
doublets mix via a mass parameter (the so called $\mu$-parameter) whose
magnitude remains to be explained. Besides, since the quartic Higgs
self-couplings are determined by the gauge couplings, the mass of the
lightest Higgs boson $h$ is constrained very stringently; in fact, the
upper limit $m_h\leq 128$ GeV has been established \cite{susyh} (the tree
level limit is $m_h\leq m_Z$, the mass of the SM neutral gauge
boson \cite{martin}).

Since at present there are not many experimental facts pointing toward
what lies beyond the SM, the best approach may be to depart from it as
little as possible. In this regard, $SU(3)_L\otimes U(1)_X$ as a flavor
group has been considered several times in the literature; first as a
family independent theory \cite{lee}, and then with a family
structure \cite{fp, long}. Some versions of the family structure provide a
solution to the problem of the number $N$ of families, in the sense that
anomaly cancellation is achieved when $N$ is a multiple of three; further,
from the condition of $SU(3)_c$ asymptotic freedom which is valid only if
the number of families is less than five, it follows that in those models
$N$ is equal to 3 \cite{fp}.

Over the last decade two three family models based on the $SU(3)_c\otimes
SU(3)_L\otimes U(1)_X$ local gauge group (hereafter the 3-3-1 structure)
have received special attention. In one of them the three known
left-handed lepton components for each family are associated to three
$SU(3)_L$ triplets \cite{fp} as $(\nu_l, l^-, l^+)_L$, where $l^+_L$ is
related to the right-handed isospin singlet of the charged lepton $l_L^-$
in the SM. In the other model the three $SU(3)_L$ lepton triplets are of
the form $(\nu_l,l^-,\nu_l^c)_L$ where $\nu_{lL}^c$ is related to the
right-handed component of the neutrino field $\nu_{lL}$ \cite{long}. In the
first model anomaly cancellation implies quarks with exotic electric
charges $-4/3$ and 5/3, while in the second one the exotic particles have
only ordinary electric charges.

All possible 3-3-1 models without exotic electric charges are presented in
Ref. \cite{ponce}, where it is shown that there are just a few anomaly
free models for one or three families, all of which have in common the
same gauge-boson content.

In this paper we are going to present the supersymmetric version of the
one-family 3-3-1 model introduced in Ref. \cite{spm}. The non-SUSY version
has the feature that the fermion states in the model are just the 27
states in the fundamental representation of the electroweak-strong
unification group $E_6$ \cite{e6}. Besides, the scale of new physics for the non-SUSY version of this model is in the range of 1-5 TeV \cite{spm, pgs}, so it is just natural to link this new scale with the SUSY scale.

Our main motivation lies in the fact that in the non-SUSY model, the three
left-handed lepton triplets and the three Higgs scalars (needed to break
the symmetry down to $SU(3)_c \otimes U(1)_Q$ in two steps) transform as
the $\bar{3}$ representation of $SU(3)_L$ and have the same quantum
numbers under the 3-3-1 structure. This becomes interesting when the
supersymmetric $N=1$ version of the model is constructed, because the
existing scalars and leptons in the model can play the role of
superpartners of each other. As a result four main consequences follow:
first, the reduction of the number of free parameters in the model as
compared to supersymmetric versions of other 3-3-1 models in the
literature \cite{susy331}; second, the result that the sneutrino, selectron
and six other sleptons do not acquire masses in the context of the model
constructed playing the role of the Goldstone bosons; third, the absence
of the $\mu$-problem, in the sense that the $\mu$-term is absent at tree
level, arising only as a result of the symmetry breaking, and fourth, the
existence of a light CP-odd scalar which may have escaped experimental
detection \cite{nir}. 

This paper is organized as follows: in Sec. \ref{sec:sec2} we briefly
review the non-supersymmetric version of the model; in Sec. \ref{sec:sec3}
we comment on its supersymmetric extension and calculate the
superpotential; in Sec. \ref{sec:sec4} we calculate the mass spectrum
(excluding the squark sector) and in Sec. \ref{sec:sec5} we present our
conclusions.

\section{\label{sec:sec2}The non-supersymmetric model} 
Let us start by describing the fermion content, the scalar sector and the
gauge boson sector of the non-supersymmetric one-family 3-3-1 model in
Ref. \cite{spm}.

We assume that the electroweak gauge group is $SU(3)_L\otimes
U(1)_X\supset SU(2)_L\otimes U(1)_Y$, that the left handed quarks (color
triplets) and left-handed leptons (color singlets) transform as the $3$
and $\bar{3}$ representations of $SU(3)_L$ respectively, that $SU(3)_c$ is
vectorlike, and that anomaly cancellation takes place family by
family as in the SM. If we begin with $Q_L^T=(u,d,D)_L$, where $(u,d)_L$
is the usual isospin doublet of quarks in the SM and $D_L$ is an isospin
singlet exotic down quark of electric charge $-1/3$, then the restriction
of having particles without exotic electric charges and the condition of
anomaly cancellation, produce the following multiplet structure for one
family \cite{spm}:

\beqa \label{campos}\nonumber
Q_{L}=\left(\ba{c} u \\ d  \\ D \ea \right)_L\sim (3,3,0), & & u^c_{L}\sim (\bar{3},1,-\frac{2}{3}), \\ \nonumber
d^c_{L}\sim (\bar{3},1,\frac{1}{3}), & &
D^c_{L}\sim (\bar{3},1,\frac{1}{3}), \\ \nonumber
L_{1L}=\left(\ba{c} e^-\\ \nu_{e}\\ N_{1}^0 \ea 
\right)_L & \sim & (1,\bar{3},-\frac{1}{3}),\\ \nonumber 
L_{2L}=\left(\ba{c} E^-
\\N_{2}^0 \\ N_{3}^0 \ea \right)_L & \sim & (1,\bar{3},
-\frac{1}{3}), \\
L_{3L}=\left(\ba{c} N^0_{4} \\ E^+ \\e^+ \ea \right)_L
& \sim & (1,\bar{3},\frac{2}{3}), 
\eeqa
where $N_1^0$ and $N_3^0$ are $SU(2)_L$ singlet exotic leptons of electric 
charge zero, and $(E^-,N_2^0)_L\cup (N_4^0,E^+)_L$ is an $SU(2)_L$ doublet 
of exotic leptons, vectorlike with respect to the SM as far as we identify 
$N_{4L}^0=N_{2L}^{0c}$. The numbers inside the parenthesis 
refer to the ($SU(3)_c, SU(3)_L, U(1)_X$) quantum numbers respectively.

In order to break the symmetry following the pattern 
\beqa \nonumber
SU(3)_c\otimes SU(3)_L\otimes U(1)_X & \longrightarrow & SU(3)_c\otimes SU(2)_L\otimes U(1)_Y  \\ 
& \longrightarrow & SU(3)_c\otimes U(1)_Q, \label{chain} 
\eeqa
and give, at the same time, masses to the fermion fields in the 
non-supersymmetric model, the following set of Higgs scalars is 
introduced \cite{spm}
\beqa \nonumber
\phi_1=\left(\ba{c}\phi^-_1 \\ \phi^0_1 \\ \phi^{'0}_1 \ea \right) & \sim
& (1,\bar{3},-\frac{1}{3}),\\ \nonumber
\phi_2=\left(\ba{c}\phi^-_2 \\ \phi^0_2
\\ \phi^{'0}_2 \ea \right) & \sim & (1,\bar{3},-\frac{1}{3}),\\
\phi_3=\left(\ba{c}\phi^0_3 \\ \phi^+_3 \\ \phi^{'+}_3 \ea \right) & \sim
& (1,\bar{3},\frac{2}{3}); \label{higgs}
\eeqa 
with Vacuum Expectation Values (VEV) given by
\beqa \label{vevs}\nonumber
&\langle\phi_1\rangle^T=(0,0,W), \qquad
\langle\phi_2\rangle^T=(0,v,0), &\\ 
& \langle\phi_3\rangle^T=(v',0,0), &
\eeqa
with the hierarchy $W>v\sim v'\sim 174$ GeV, the electroweak breaking
scale. From Eqs. (\ref{campos}) and (\ref{higgs}) we can see that the
three left-handed lepton triplets and the three Higgs scalars have the
same quantum numbers under the 3-3-1 gauge group, so they can play the
role of superpartners. Also, the isospin doublet in $\phi_2$ plays the role
of $\phi_d$ and the isospin doublet in $\phi_3$ plays the role of $\phi_u$
in extensions of the SM with two Higgs doublets (2HDM), in which $\phi_d$ 
couples only to down type quarks and $\phi_u$ couples only to up type 
quarks (2HDM Type II).

There are a total of 17 gauge bosons in this 3-3-1 model. One gauge field
$B^\mu$ associated with $U(1)_X$, the 8 gluon fields $G^\mu$ associated
with $SU(3)_c$ which remain massless after breaking the symmetry, and
another 8 gauge fields $A^\mu$ associated with $SU(3)_L$ and that we write 
for convenience in the following way:

\[{1\over 2}\lambda_\alpha A^\mu_\alpha={1\over \sqrt{2}}\left(
\begin{array}{ccc}D^\mu_1 & W^{+\mu} & K^{+\mu} \\ W^{-\mu} & D^\mu_2 &
K^{0\mu} \\ K^{-\mu} & \bar{K}^{0\mu} & D^\mu_3 \end{array}\right), \]
where $D^\mu_1=A_3^\mu/\sqrt{2}+A_8^\mu/\sqrt{6},\;
D^\mu_2=-A_3^\mu/\sqrt{2}+A_8^\mu/\sqrt{6}$, and
$D^\mu_3=-2A_8^\mu/\sqrt{6}$. $\lambda_i, \; i=1,2,...,8$ are the eight
Gell-Mann matrices normalized as $Tr(\lambda_i\lambda_j)  
=2\delta_{ij}$.

The covariant derivative for this model is given by the expression:
$D^\mu=\partial^\mu-
i(g_3/2)\lambda^\alpha G_\mu^\alpha - 
i(g_2/2)\lambda^\alpha A_\mu^\alpha - ig_1XB^\mu$,
where $g_i,\; i=1,2,3$ are the gauge coupling constants for $U(1)_X, \; 
SU(3)_L$ and $SU(3)_c$ respectively.

The sine of the electroweak mixing angle is given by 
$S_W^2=3g_1^2/(3g_2^2+4g_1^2)$. The photon field is thus:
\begin{equation}\label{foton}
A_0^\mu=S_WA_3^\mu+C_W\left[\frac{T_W}{\sqrt{3}}A_8^\mu + 
\sqrt{(1-T_W/3)}B^\mu\right],
\end{equation}
where $C_W$ and $T_W$ are the cosine and tangent of the electroweak mixing 
angle. 

Finally, the two neutral currents in the model are defined as:
\begin{eqnarray}\nonumber
Z_0^\mu&=&C_WA_3^\mu-S_W\left[\frac{T_W}{\sqrt{3}}A_8^\mu + 
\sqrt{(1-T_W/3)}B^\mu\right], \\ \label{zetas}
Z_0^{\prime\mu}&=&-\sqrt{(1-T_W/3)}A_8^\mu+\frac{T_W}{\sqrt{3}}B^\mu,
\end{eqnarray}
where $Z^\mu$ coincides with the weak neutral current of the SM, with the 
gauge boson associated with the $Y$ hypercharge given by:
\[Y^\mu=\left[\frac{T_W}{\sqrt{3}}A_8^\mu + 
\sqrt{(1-T_W/3)}B^\mu\right]. \]

The consistency of the model requires the existence of eight Goldstone
bosons in the scalar spectrum, out of which four are charged and four are
neutral (one CP-even state and three CP-odd) \cite{pgs} in order to
provide with masses for $W^\pm, \; K^\pm ,\; K^0, \; \bar{K}^0, \; Z^0$
and $Z^{\prime 0}$.

\section{\label{sec:sec3}The Supersymmetric extension} 
When we introduce supersymmetry in the SM, the entire spectrum of
particles is doubled as we must introduce the superpartners of the known
fields, besides two scalar doublets $\phi_u$ and $\phi_d$ must be used in
order to cancel the triangle anomalies; then the superfields
${\hat\phi}_u$, and ${\hat\phi}_d$, related to the two scalars, may couple
via a term of the form $\mu {\hat\phi}_u{\hat\phi}_d$ which is gauge and
supersymmetric invariant, and thus the natural value for $\mu$ is expected
to be much larger than the electroweak and supersymmetry breaking scales.
This is the so-called $\mu$ problem.

However, in a non supersymmetric model as the one presented in the former
section, in which the Higgs fields and the lepton fields transform
identically under the symmetry group, we can have (as far as we take
proper care of the mass generation and the symmetry breaking pattern) the
three lepton triplets and the three Higgs triplets as the superpartners of
each other. Consequently, we can construct the supersymmetric version of
our model without the introduction of Higgsinos, with the supersymmetric
extension automatically free of chiral anomalies.

For one family we thus end up with the following seven chiral superfields:
$\hat{Q}$, $\hat{u}$, $\hat{d}$, $\hat{D}$, $\hat{L}_1$, $\hat{L}_2$, and
$\hat{L}_3$, plus gauge bosons and gauginos. The identification of the 
gauge bosons eigenstates in the SUSY version follows the non-SUSY analysis 
as we will show below.
 
\subsection{\label{sec:sub31}The Superpotential}
Let us now write the most general $SU(3)_c\otimes SU(3)_L\otimes U(1)_X$ 
invariant superpotential
\beqa \nonumber 
U&=& \sum_a(h^u\hat{Q}_a\hat{u}\hat{L}^a_3 + 
\lambda^{(1)}\hat{Q}_a\hat{d}\hat{L}^a_1 +
h^d\hat{Q}_a\hat{d}\hat{L}^a_2 \\ \nonumber &+& \lambda^{(2)}\hat{Q}_a\hat{D}\hat{L}^a_1 
+h^D\hat{Q}_a\hat{D}\hat{L}^a_2) + \lambda^{(3)}\hat{u}\hat{d}\hat{D} \\ 
&+& \sum_{abc} \epsilon_{abc}(h^e\hat{L}^a_1\hat{L}^b_2\hat{L}^c_3 
+ \lambda^{(4)}\hat{Q}^a\hat{Q}^b\hat{Q}^c),
\label{superpotential} 
\eeqa 
\noindent
where $a,b,c=1,2,3$ are $SU(3)_L$ tensor indices and the chirality 
and color indices have been
omitted.  Notice the absence of terms bilinear in the superfields, so a
bare $\mu$ term is absent in the superpotential $U$, but it can be
generated, after symmetry breaking, by one of the terms in
Eq.(\ref{superpotential}); as a matter of fact it is proportional to
$h^e(\langle\tilde{N}_1^0 \rangle\tilde{N}^0_2 + \langle\tilde{N}_3^0\rangle\tilde{\nu})\tilde{N}^0_4$, where $\langle...\rangle$ stands for the VEV of the neutral scalar field inside the brackets and the tilde denotes the superpartner of the respective field. This effective $\mu$ term is at most of the order of the supersymmetry breaking scale, but as we will show
in the next section $h^e \approx 0$ in order to
have a consistent supersymmetric model.  This is the way how the
Supersymmetric $\mu$ problem is avoided in the context of the model in
this paper. 

The $\hat{u}\hat{d}\hat{D}$ and $\hat{Q}\hat{Q}\hat{Q}$ terms violate
baryon-number and can possibly lead to rapid proton decay. We may forbid
these interactions by introducing the following baryon-parity
\begin{eqnarray}\nonumber
(\hat{Q}, \hat{u}, \hat{d}, \hat{D}) & \rightarrow & -(\hat{Q}, 
\hat{u}, \hat{d}, \hat{D}), \\ 
(\hat{L}_1, \hat{L}_2, \hat{L}_3)
& \rightarrow & + (\hat{L}_1, \hat{L}_2, \hat{L}_3).
\end{eqnarray}

This protects the model from too fast proton decay, but the superpotential 
still contains operators inducing lepton number violation. This is desirable 
if we want to describe Majorana masses for the neutrinos in our model.

Another discrete symmetry worth considering is $L_{1L}\leftrightarrow
L_{2L}$, which implies $h^e=0$, $\lambda^{(1)}=h^d$ and
$\lambda^{(2)}=h^D$. As we will see in the next section, a very small value
of $h^e$ is mandatory for having a neutrino with a very small 
tree-level mass.

\subsection{\label{sec:sub32}The scalar potential} 
The scalar potential is written as 
\begin{equation} \label{potesc}
V_{SP}=V_F+V_D+V_{\hbox{soft}},
\end{equation} 
where the first two terms come from the exact SUSY sector, while the last
one is the sector of the theory that breaks SUSY explicitly.

We now display the different terms in Eq. (\ref{potesc}):
\beqa \nonumber
V_F&=&\sum_i\left |\frac{\partial U}{\partial\phi_i}
\right|^2 \\ \nonumber
&=& |h^e|^2(|\tilde{L}_1|^2|\tilde{L}_2|^2+
|\tilde{L}_1|^2|\tilde{L}_3|^2+|\tilde{L}_2|^2
|\tilde{L}_3|^2 \\ 
& & -|\tilde{L}_1^\dag \tilde{L}_2|^2-
|\tilde{L_1}^\dagger \tilde{L}_3|^2-|\tilde{L}_2^\dagger 
\tilde{L}_3|^2).
\eeqa

\beqa V_D=\frac{1}{2}D^\alpha D^\alpha+
\frac{1}{2}D^2\,,\nonumber\eeqa
where
\beqa \nonumber
D^\alpha= g_2\sum_{i=1}^3\sum_{a,b=1}^8 L_{i,a}^*
(\frac{{-\lambda^\alpha}^*}{2})_{ab} L_{i,b} \quad 
(\alpha=1,..., 8),
\eeqa
and
\beqa \nonumber
D=g_1 \sum_{i=1}^3 \sum_{a=1}^8 L_{i,a}^*X(L) L_{i,a}\,
\eeqa
\noindent 
($a,b=1,2,\dots 8$ are $SU(3)_L$ tensor indices). Then we have
\begin{widetext}
\beqa V_D&=&\frac{1}{2}g_2^2\big[\frac{1}{3}\big 
\{(\tilde{L}_1^{\dagger} \tilde{L}_1)^2+(\tilde{L}_2^{\dagger} 
\tilde{L}_2)^2+(\tilde{L}_3^{\dagger} 
\tilde{L}_3)^2-(\tilde{L}_1^{\dagger} \tilde{L}_1)
(\tilde{L}_2^{\dagger} \tilde{L}_2)\nonumber\\
& &-\,(\tilde{L}_1^{\dagger} \tilde{L}_1)(\tilde{L}_3^{\dagger}
 \tilde{L}_3)
-(\tilde{L}_2^{\dagger} \tilde{L}_2)(\tilde{L}_3^{\dagger} 
\tilde{L}_3)\big\}+|\tilde{L}_1^\dag \tilde{L}_2|^2+
|\tilde{L}_1^\dag \tilde{L}_3|^2+|\tilde{L}_2^\dag 
\tilde{L}_3|^2\big ]\nonumber\\& &+\frac{1}{18}g_1^2
\big[\big (\tilde{L}_1^{\dagger} \tilde{L}_1)^2+
(\tilde{L}_2^{\dagger} \tilde{L}_2)^2+4
(\tilde{L}_3^{\dagger} \tilde{L}_3)^2+2(
\tilde{L}_1^{\dagger} \tilde{L}_1)(\tilde{L}_2^{\dagger}
 \tilde{L}_2)\nonumber\\& &-\,4(\tilde{L}_1^{\dagger} 
\tilde{L}_1)(\tilde{L}_3^{\dagger} \tilde{L}_3)
-4(\tilde{L}_2^{\dagger} \tilde{L}_2)
(\tilde{L}_3^{\dagger} \tilde{L}_3)\big ] 
\eeqa
\end{widetext}
(On deriving $V_F$ and $V_D$ we have used the identities
$\epsilon_{ijk}\epsilon^{ilm}=\delta_j^l
\delta_k^m-\delta_j^m\delta_k^l$, and 
$\lambda^\alpha_{ij}
\lambda^\alpha_{kl}=2\delta_{il}\delta_{jk}- 
\frac{2}{3}\delta_{ij}\delta_{kl}).$

Finally, the soft SUSY-breaking potential is given by  
\beqa \nonumber
V_{\hbox{soft}}&=&
m_{L_1}^2\tilde{L}_1^{\dagger} \tilde{L}_1+m_{L_2}^2
\tilde{L}_2^{\dagger}\tilde{L}_2+m_{L_3}^2
\tilde{L}_3^{\dagger} \tilde{L}_3 \\ \nonumber
&+& m_{L_1L_2}^2\hbox{Re}\,(\tilde{L}_1^{\dagger} \tilde{L}_2)
+ h^\prime\hbox{Re}\,(\epsilon^{abc}\tilde{L}_a
\tilde{L}_b\tilde{L}_c) \\ 
&+& \frac{M_1}{2}\tilde{B}^0\tilde{B}^0 + 
\frac{M_2}{2}\sum_{a=1}^8 \tilde{A}_a\tilde{A}^a + \dots ,
\eeqa
where $M_1$ is the soft mass parameter of the $U(1)_X$ gaugino and $M_2$ refers to the soft mass parameter of the $SU(3)_L$ gauginos. 

After redefining
$(\tilde{E}^-,\tilde{N}_2)$ as $\phi_d$ and $(\tilde{N}_4,\tilde{E}^+)$ as
$\phi_u$, the parts of $V=V_F+V_D$ containing the sleptons are given by
\begin{widetext}
\beqa\label{VV} V&=&\delta 
[(\phi_d^\dagger 
\phi_d+\tilde{N}_3^\dagger\tilde{N}_3)^2+(\tilde{e}^+\tilde{e}^- + 
\tilde{\nu}^\dagger \tilde{\nu} + \tilde{N}_1^\dagger\tilde{N}_1)^2] + 
\eta (\phi_u^\dagger \phi_u + \tilde{e}^+ \tilde{e}^-)^2 
\nonumber \\
& &+\gamma (\phi_d^\dagger \phi_u + 
\tilde{N}_3^\dagger\tilde{e}^+)(\phi_u^\dagger 
\phi_d + \tilde{e}^-\tilde{N}_3) + \beta 
(\phi_d^\dagger \phi_d + 
\tilde{N}_3^\dagger\tilde{N}_3^+)(\phi_u^\dagger 
\phi_u + \tilde{e}^+\tilde{e}^-) \nonumber \\
& & +\alpha (\phi_d^\dagger \phi_d + 
\tilde{N}_3^\dagger\tilde{N}_3)(|\tilde{e}|^2+|\tilde{\nu}|^2 + 
|\tilde{N}_1|^2) \nonumber \\
& & +\beta  (\phi_u^\dagger \phi_u + 
\tilde{e}^+\tilde{e}^-)(|\tilde{e}|^2+|\tilde{\nu}|^2 + 
|\tilde{N}_1|^2) \nonumber \\
& & + \gamma 
(\tilde{e}^+\tilde{E}^- + \tilde{\nu}^\dagger\tilde{N}_2 + 
\tilde{N}_1^\dagger\tilde{N}_3)
(\tilde{E}^+\tilde{e}^- + \tilde{N}_2^\dagger\tilde{\nu}_2 + 
\tilde{N}_3^\dagger\tilde{N}_1)\nonumber \\
& & + \gamma (\tilde{e}^+\tilde{N}_4^- + \tilde{\nu}^\dagger\tilde{E}^+ + 
\tilde{N}_1^\dagger\tilde{e}^+) (\tilde{N}_4^\dagger\tilde{e}^- + 
\tilde{E}^-\tilde{\nu} + \tilde{e}^-\tilde{N}_1),
\eeqa
\end{widetext}
where $\delta = \left(\frac{g_2^2}{6}+\frac{g_1^2}{18}\right)$, 
$\eta = \left(\frac{g_2^2}{6}+\frac{2g_1^2}{9}\right)$, 
$\gamma = \left(\frac{g_2^2}{2}-|h^e|^2\right)$, 
$\beta = \left(|h^e|^2-\frac{g_2^2}{6}-\frac{2g_1^2}{9}\right)$, and  
$\alpha = \left(|h^e|^2-\frac{g_2^2}{6}+\frac{g_1^2}{9}\right)$.

\section{\label{sec:sec4}Mass Spectrum}
Masses for the particles are generated in this model from the VEV of the
scalar fields and from the soft terms in the superpotential.

For simplicity we assume that the VEVs are real, which means 
that spontaneous CP violation through the scalar exchange is not 
considered in this work. Now, for convenience in reading we rewrite the 
expansion of the scalar fields acquiring VEVs as:
\begin{eqnarray} \nonumber \label{neutras}
\tilde{N}_1^0&=&\langle\tilde{N}_1^0\rangle 
+ \frac{\tilde{N}_{1R}^0+i\tilde{N}_{1I}^0}{\sqrt{2}}, \\ \nonumber
\tilde{N}_2^0&=&\langle\tilde{N}_2^0\rangle 
+ \frac{\tilde{N}_{2R}^0+i\tilde{N}_{2I}^0}{\sqrt{2}}, \\ \nonumber
\tilde{N}_3^0&=&\langle\tilde{N}_3^0\rangle 
+ \frac{\tilde{N}_{3R}^0+i\tilde{N}_{3I}^0}{\sqrt{2}}, \\ \nonumber
\tilde{N}_4^0&=&\langle\tilde{N}_4^0\rangle 
+ \frac{\tilde{N}_{4R}^0+i\tilde{N}_{4I}^0}{\sqrt{2}}, \\
\tilde{\nu}_e&=&\langle\tilde{\nu}\rangle 
+ \frac{\tilde{\nu}_R+i\tilde{\nu}_I}{\sqrt{2}}, 
\end{eqnarray}
in an obvious notation taken from Eq.(\ref{campos}), 
where $\tilde{N}_{iR}^0\;(\tilde{\nu}_R)$ and
$\tilde{N}_{iI}^0\;(\tilde{\nu}_I)\;\; i=1,2,3,4$ refer, respectively, to 
the real sector and to the imaginary sector of the sleptons. 
In general, $\langle\tilde{\nu}_e\rangle$ 
and $\langle\tilde{N}_i^0\rangle ,\; i=1,2,3,4$ can be all different
from zero, but as we will see in the following analysis there are some 
constraints relating them. Also 
$\langle\tilde{\nu}\rangle \leq 0.2$ TeV, $\langle\tilde{N}_i\rangle\leq
0.2$ TeV, for $i=2,4$ in order to respect the SM phenomenology, and
$\langle\tilde{N}_j\rangle\geq 1$ TeV, for $j=1,3$ in order to respect
the phenomenology of the 3-3-1 model in Refs. \cite{spm, pgs}.

Our approach will be to look for consistency in the sense that the mass
spectrum must include a light spin 1/2 neutral particle (the neutrino)
with the other spin 1/2 neutral particles having masses larger than or
equal to half of the $Z^0$ mass, to be in agreement with experimental
bounds. Also we need eight spin zero Goldstone bosons, four charged and
four neutral ones, out of which one neutral must be related to the real
sector of the sleptons and three neutrals to the imaginary sector, in
order to produce masses for the gauge bosons after the breaking of the
symmetry.

As we will show in this section, a consistent set of VEV is provided by
$\langle\tilde{\nu}_e\rangle = v$, $\langle\tilde{N}_3^0\rangle =V, \;  
\langle\tilde{N}_2^0\rangle =v_d$ and $\langle\tilde{N}_4^0\rangle =v_u$,
with the hierarchy $V >v_u\sim v_d\sim v$, and the constraint
$\langle\tilde{N}_1\rangle=-vv_d/V$.  This situation implies a symmetry
breaking pattern of the form $SU(3)_c\otimes SU(3)_L\otimes U(1)_X
\longrightarrow SU(3)_c\otimes U(1)_Q$, instead of the chain in
Eq.(\ref{chain}). So, we can not claim that the MSSM is an effective
theory of the model presented here; rather the model here is an
alternative to the MSSM so well analyzed in the literature \cite{mssm,
susyh, martin}.

Playing with the VEV and the other parameters in the superpotential, 
special attention must be paid to the several constraints coming from the 
minimization of the scalar potential, which at tree level are:
\beqa \nonumber
\langle\tilde{N}_2^0\rangle\langle\tilde{\nu}\rangle&=& 
- \langle\tilde{N}_3^0\rangle\langle\tilde{N}_1\rangle ,\\
m^2_{L_1L_2}&=& h^\prime \langle\tilde{N}_4^0\rangle
\frac{\langle\tilde{N}_3^0\rangle \langle\tilde{N}_1^0\rangle + 
\langle\tilde{\nu}\rangle\langle\tilde{N}_2^0\rangle}
{\langle\tilde{N}_3^0\rangle \langle\tilde{\nu}\rangle
-\langle\tilde{N}_2^0\rangle \langle\tilde{N}_1^0\rangle}=0, \nonumber  \\
m^2_{L_1}&=&-\alpha(\langle\tilde{N}_2^0\rangle^2 + 
\langle\tilde{N}_3^0\rangle^2)-\beta\langle\tilde{N}_4^0\rangle^2
-2\delta (\langle\tilde{\nu}\rangle^2  \nonumber \\ 
& & + \langle\tilde{N}_1^0\rangle^2)-h^\prime\frac{\langle\tilde{N}_4^0\rangle 
\langle\tilde{N}_3^0\rangle}
{2\langle\tilde{\nu}\rangle}, \nonumber \\
m^2_{L_2}&=& -\beta\langle\tilde{N}^0_4\rangle^2-\alpha 
(\langle\tilde{\nu}\rangle^2+\langle\tilde{N}_1^0\rangle^2) 
-2\delta(\langle\tilde{N}_2^0\rangle^2\nonumber \\
& & +\langle\tilde{N}_3^0\rangle^2) -h^\prime\frac{\langle\tilde{N}_4^0\rangle 
\langle\tilde{N}_1^0\rangle}
{2\langle\tilde{N}^0_2\rangle}, \nonumber \\
m^2_{L_3}&=& -\beta ( \langle\tilde{\nu}\rangle^2 + 
 \langle\tilde{N}_1^0\rangle^2 + \langle\tilde{N}_2^0\rangle^2+
\langle\tilde{N}_3^0\rangle^2)  \nonumber \\  
& & - 2\eta \langle\tilde{N}_4^0\rangle^2-h^\prime
\frac{\langle\tilde{N}_3^0\rangle \langle\tilde{\nu}\rangle - 
\langle\tilde{N}_1^0\rangle  \langle\tilde{N}_2^0\rangle}
{2 \langle\tilde{N}_4^0\rangle}, \nonumber \\ \label{minim} 
\eeqa
where $\alpha ,\beta ,\delta$ and $\eta$ were defined above. The result 
$m_{L_1L_2}=0$ comes from the first constraint and has important 
consequences as we will see in what follows.

\subsection{\label{sec:sub41}Spectrum in the Gauge Boson Sector}
With the most general VEV structure presented in Eq. (\ref{neutras}), the 
charged gauge bosons $W^\pm_\mu$ and $K^\pm_\mu$ mix up and the 
diagonalization of the corresponding squared-mass matrix yields the 
masses \cite{pgs}
\begin{eqnarray}\nonumber
M_{W'}^2&=&{g_2^2 \over 2}(\langle\tilde{N}_4^0\rangle^2+\langle\tilde{\nu}\rangle^2 + 
\langle\tilde{N}^0_1\rangle^2), \\ 
M_{K'}^2&=&{g_2^2 \over 2}(\langle\tilde{N}_3^0\rangle^2+\langle\tilde{N}_4^0\rangle^2+
\langle\tilde{N}_2^0\rangle^2),
\end{eqnarray}
related to the physical fields $W^{'}_\mu=\eta (\langle\tilde{N}_2^0\rangle K_\mu-\langle\tilde{N}_3^0\rangle W_\mu)$ and $K^{'}_\mu = \eta(\langle\tilde{N}_3^0\rangle K_\mu+\langle\tilde{N}_2^0\rangle W_\mu)$ associated with the known charged current $W^{'\pm}_\mu$, and the new one $K^{'\pm}_\mu$ predicted in the context of this model ($\eta^{-2}=\langle\tilde{N}_2^0\rangle^2+\langle\tilde{N}_3^0\rangle^2$ is a normalization factor).
Notice that with the hierarchy $\langle\tilde{N}_3^0\rangle >> \langle\tilde{N}_2^0\rangle \sim \langle\tilde{N}_4^0\rangle \sim \langle\tilde{\nu}\rangle $, the mixing between $W^\pm_\mu$ and $K^\pm_\mu$ is well under control due to fact that the physical 
$W^{\prime\pm}$ is mainly the $W^\pm$ of the weak basis, with a small 
component along $K^\pm$ of the order of $\langle\tilde{N}_2^0\rangle/\langle\tilde{N}_3^0\rangle$. 

The expression for the $W^{\prime\pm}$ mass combined with the minimization
conditions in Eq.({\ref{minim}}) implies
$(\langle\tilde{N}_4^0\rangle^2+\langle\tilde{\nu}\rangle^2 +
\langle\tilde{\nu}\rangle^2\langle\tilde{N}^0_2\rangle^2/
\langle\tilde{N}^0_3\rangle^2)^{1/2}\approx 174$ GeV.

For the five electrically neutral gauge bosons we get first, that the
imaginary part of $K^0_\mu$ decouples from the other four electrically
neutral gauge bosons, acquiring a mass
$M^2_{K^0_I}=(g_2^2/4)(\langle\tilde{\nu}\rangle^2
+\langle\tilde{N}_2^0\rangle^2 +\langle\tilde{N}_3^0\rangle^2
+\langle\tilde{N}_4^0\rangle^2)$ \cite{pgs}. Now, in the basis $(B^\mu,
A^\mu_3, A^\mu_8,K^{0\mu}_R)$, the obtained squared-mass matrix has
determinant equal to zero which implies that there is a zero eigenvalue
associated to the photon field with eigenvector $A_0^\mu$ as given in Eq.
(\ref{foton}).

The mass matrix for the neutral gauge boson sector can now be written in
the basis $(Z_0^{'\mu}, Z_0^\mu, K^{0\mu}_R)$, where the fields
$Z_0^{'\mu}$ and $Z_0^\mu$ have been defined in Eqs. (\ref{zetas}). We can
diagonalize this mass matrix in order to obtain the physical fields, but
the mathematical results are not very illuminating. Since
$\langle\tilde{N}_3^0\rangle >> \langle\tilde{N}_2^0\rangle \sim
\langle\tilde{N}_4^0\rangle \sim \langle\tilde{\nu}\rangle \sim 174$ GeV,
we perform a perturbation analysis for the particular case
$\langle\tilde{N}_2^0\rangle = \langle\tilde{N}_4^0\rangle =
\langle\tilde{\nu}\rangle \equiv v$ using
$q=v/\langle\tilde{N}_3^0\rangle$ as the expansion parameter. In this way
we obtain one eigenvalue of the form 
\begin{equation} M^2_{Z_1}\approx
g_2^2C_W^{-2}v^2 \left(1+{1\over 8}q^2(7 + 6 T_W^2 - 9 T_W^4)\right), 
\end{equation} 
and other two of the order $\langle\tilde{N}_3^0\rangle^2$ \cite{pgs}. So,
we have a neutral current associated to a mass scale $v\simeq 174$ GeV
which may be identified with the known SM neutral current, and two new
electrically neutral currents associated to a mass scale
$\langle\tilde{N}_3^0\rangle>>v$.

Now, using the expressions for $M_{W'}$ and $M_{Z_1}$ we obtain for the
$\rho$ parameter at tree-level\cite{hunter} 
\begin{equation}
\rho=M^2_{W'}/(M^2_{Z_1}C^2_W) \simeq 1-{3 \over 8}q^2(1 + 2 T_W^2 - 3
T_W^4), \end{equation} 
so that the global fit $\rho=1.0012^{+0.0023}_{-0.0014}$ \cite{langa}
provides us with the lower limit $\langle\tilde{N}_3^0\rangle \geq 8.7$
TeV (where we are using $S^2_W=0.23113$ \cite{pdb} and neglecting loop
corrections which depend on the splitting of the $SU(2)_L$ doublets).

This result justifies both the imposition of the hierarchy
$\langle\tilde{N}_3^0\rangle >> \langle\tilde{N}_2^0\rangle \sim
\langle\tilde{N}_4^0\rangle \sim \langle\tilde{\nu}\rangle$ and the
existence of the expansion parameter $q \leq 0.02$. This in turn shows,
first that the small component
$\left(\langle\tilde{N}_2^0\rangle/\sqrt{\langle\tilde{N}_2^0\rangle^2
+\langle\tilde{N}_3^0\rangle^2}\right)K_\mu$ of the eigenstate
$W^{\prime}_\mu$ will contaminate tree-level physical processes at most at
the level of 2\% (by the way, such a mixing can contribute to the $\Delta
I=1/2$ enhancement in nonleptonic weak processes), and second that the
estimated order of the masses of the new charged and neutral gauge bosons
in the model are not in conflict neither with constraints on their mass
scale calculated from a global fit of data relevant to electron-quark
contact interactions \cite{barger}, nor with the bounds obtained in 
$p\bar{p}$ collisions at the Tevatron \cite{abe}.

\subsection{\label{sec:sub42}Masses for the Quark Sector}
Let us assume in the following analysis that we are working with the 
third family. The first term in the superpotential produces for the up 
type quark a mass $m_t=h^u \langle \tilde{N}^0_4 \rangle = 174$ GeV, which 
implies $\langle\tilde{N}_4^0\rangle\approx 10^2$ 
GeV and $h^u\sim 1$, while for the down type quarks the second to
fifth terms generate, in the basis $(d,D)\; [(d_R,D_R)$ column and 
$(d_L,D_L)$ row], the mass matrix 
\begin{equation}\nonumber 
M_{dD}=\left(\begin{array}{cc}\lambda^{(1)}\langle \tilde{\nu} \rangle
+ h^d\langle \tilde{N}^0_2 \rangle & \lambda^{(1)}\langle\tilde{N}^0_1 
\rangle + h^d\langle \tilde{N}^0_3\rangle \\ 
\lambda^{(2)}\langle 
\tilde{\nu}\rangle + h^D\langle \tilde{N}^0_2 \rangle  & 
 \lambda^{(2)}\langle \tilde{N}^0_1 
\rangle + h^D\langle \tilde{N}^0_3 \rangle  \end{array}  \right),
\end{equation} 
which produces a mass of the order of $\langle\tilde{N}_3^0\rangle$ for 
the exotic quark $D$, and for the ordinary quark $d$ a mass of the order 
of $(\langle\tilde{\nu}\rangle+\langle\tilde{N}^0_2\rangle)$, suppressed 
by differences of Yukawa couplings (it is zero for $\lambda^{(1)}=h^d$ 
and $\lambda^{(2)}=h^D$).

Using the former results and the expression for the $W^\pm$ mass it 
follows that
$\langle\tilde{N}_4^0\rangle\approx\langle\tilde{\nu}\rangle\approx
\langle\tilde{N}_2^0\rangle\approx 10^2$ GeV.

It is worth noticing that the isospin doublet in $\tilde{L}_{3L}$ couples
only to up type quarks, while the isospin doublets in $\tilde{L}_{1L}$ and
$\tilde{L}_{2L}$ couple only to down type quarks.

\subsection{\label{sec:sub43}Masses for neutralinos}
The neutralinos are linear combinations of neutral gauginos and neutral
leptons (there are not Higgsinos). For this model and in 
the basis $(\nu_e, N_1, N_2, N_3, N_4,
\tilde{B}^0, \tilde{A}_3, \tilde{A}_8, \tilde{K}^0, \tilde{\bar{K}}^0)$, 
their mass matrix is given by

\begin{equation}
M_{ntns}= \left(\begin{array}{cc} M_N & M^T_{gN} \\ 
M_{gN} & M_g  \end{array} \right),
\end{equation}
where $M_N$ is the matrix 
\begin{equation} \label{masslepneu}
M_N={h^e \over 2 }\left(\begin{array}{ccccc}
0 & 0 & 0 & \langle \tilde{N}^0_4 \rangle & 
\langle \tilde{N}^0_3 \rangle \\
0 & 0 & -\langle \tilde{N}^0_4 \rangle & 0 & 
-\langle \tilde{N}^0_2 \rangle \\
0 & -\langle \tilde{N}^0_4 \rangle & 0 & 0 & 
-\langle \tilde{N}^0_1 \rangle \\
\langle \tilde{N}^0_4 \rangle & 0 & 0 & 0 & 
\langle \tilde{\nu} \rangle \\
\langle \tilde{N}^0_3 \rangle & 
-\langle \tilde{N}^0_2 \rangle & 
-\langle \tilde{N}^0_1 \rangle & 
\langle \tilde{\nu} \rangle & 0 
\end{array}\right),
\end{equation}
\noindent
$M_{gN}$ is given by 
\begin{widetext}
\begin{equation} 
M_{gN}=\left(\begin{array}{ccccc}
-g_1{\sqrt{2}\over 3} \langle \tilde{\nu} \rangle & 
-g_1{\sqrt{2}\over 3} \langle \tilde{N}^0_1 \rangle & 
-g_1{\sqrt{2}\over 3} \langle \tilde{N}^0_2 \rangle & 
-g_1{\sqrt{2}\over 3} \langle \tilde{N}^0_3 \rangle & 
g_1 {2\sqrt{2}\over 3} \langle \tilde{N}^0_4 \rangle \\
g_2{1\over \sqrt{2}} \langle \tilde{\nu} \rangle  & 
0 & g_2{1\over \sqrt{2}} \langle \tilde{N}_2^0 \rangle  & 
0 &
-g_2{1\over \sqrt{2}} \langle \tilde{N}^0_4 \rangle \\
-g_2{1\over \sqrt{6}} \langle \tilde{\nu} \rangle & 
g_2{2\over \sqrt{6}} \langle \tilde{N}^0_1 \rangle & 
-g_2{1\over \sqrt{6}} \langle \tilde{N}^0_2 \rangle & 
g_2{2\over \sqrt{6}} \langle \tilde{N}^0_3 \rangle & 
-g_2{1\over \sqrt{6}} \langle \tilde{N}^0_4 \rangle \\
-g_2\langle \tilde{N}^0_1 \rangle & 0 & 
-g_2\langle \tilde{N}^0_3 \rangle & 0 & 0 \\
0 & -g_2\langle \tilde{\nu} \rangle & 0 & 
-g_2\langle \tilde{N}^0_2 \rangle & 0 
\end{array}\right), 
\end{equation}
\end{widetext}
and from the soft terms in the superpotential we read $M_g = Diag(M_1,
M_2, M_2, A_{2\times 2})$, where $A_{2\times 2}$ is a $2\times 2$ matrix
with entries zero in the main diagonal and $M_2$ in the secondary 
diagonal.

Now, in order to have a consistent model, one of the eigenvalues of this
mass matrix must be very small (corresponding to the neutrino field), with
the other eigenvalues larger than half of the $Z^0$ mass. It is clear that
for $h^e$ very small and simultaneously $M_i,\; i=1,2$ very large, we have
a see-saw type mass matrix; but $M_i,\; i=1,2$ very large is inconvenient
because it restores the hierarchy problem.

A detailed analysis shows that $M_{ntns}$ contains two Dirac neutrinos and
six Majorana neutral fields, and that for $M_i\leq 10$ TeV, $i=1,2$ we
have a mass spectrum consistent with the low energy phenomenology only if
$h^e\approx 0$. By imposing $h^e=0$, a zero tree-level Majorana mass for
the neutrino is obtained, with the hope that the radiative corrections
should produce a small mass. (The symmetry $L_{1L}\leftrightarrow L_{2L}$
implies $h^e=0$).

To diagonalize $M_{ntns}$ analytically is a hopeless task, so we propose a
controlled numerical analysis using fixed values for some parameters as
suggested by the low energy phenomenology (for example $g_1$(TeV)$\approx
0.38$ and $g_2$(TeV)$\approx 0.65$) and leaving free other parameters, but
in a range of values bounded by theoretical and experimental restrictions.
With this in mind we use 0.1 TeV $\leq M_i\leq$ 10 TeV, $i=1,2$ (in order
to avoid the hierarchy problem) and $h^e\approx 0$ (in order to have a
consistent mass spectrum).

The random numerical analysis with the constraints stated above shows that
for $M_1\approx 0.35$ TeV, $M_2\approx 3.1$ TeV,
$\langle\tilde{N}_3^0\rangle\approx 10 $ TeV, $\langle\tilde{N}_4^0\rangle
\approx 150$ GeV, $\langle\tilde{N}_2^0\rangle = \langle\tilde{\nu}\rangle
\approx 80$ GeV, $\langle\tilde{N}_1^0\rangle$ calculated from the
constraints coming from the minimum of the scalar potential (see Eq.
(\ref{minim})), and $h^e\approx 0$, we get a neutrino mass of a few
electron volts, while all the other neutral fields acquire masses above 45
GeV as desired. Also, the analysis is quite insensitive to the variation
of the parameters, with the peculiarity that an increase in $M_1$ and
$M_2$ implies an increase in $\langle\tilde{N}_3^0\rangle$.

We are going to use from now on the notation $\langle\tilde{\nu}\rangle=v,
\; \langle\tilde{N}_4\rangle=v_u, \;  \langle\tilde{N}_2^0\rangle = v_d,
\; \langle\tilde{N}_3^0\rangle = V$, with $\langle\tilde{N}_1^0\rangle =
-v_dv/V$ as constrained by the minimization conditions in Eq.
(\ref{minim}).

Another possibility with $h^e\neq 0$ but very small demands for
$\langle\tilde{\nu}\rangle = \langle\tilde{N}_1^0\rangle = 0$, and
produces a lightest neutralino only in the KeV scale, which may be
adequate for the second and third family, but not for the first one. The
advantage of this particular case is that it reduces to the study of the
scalar potential presented in Ref.\cite{pgs} for the non-supersymmetric
case, with an analysis of the mass spectrum similar to the one in that
paper.

\subsection{\label{sec:sub44}Masses for the scalar sector}
For the scalars we have three sectors, one charged and two neutrals (one 
real and the other one imaginary) which do not mix, so we can consider 
them separately.

\subsubsection{\label{sec:subsub441}The charged scalars sector}
For the charged scalars in the basis $(\tilde{e}_1^-, \tilde{e}_2^-, 
\tilde{E}_1^-, \tilde{E}_2^-)$, we get the squared-mass matrix:
\begin{widetext}
\[\left(\begin{array}{cccc} 
2\gamma v_u^2-h^\prime v_uV/v & 2\gamma \langle\tilde{N}^0_1\rangle v_u + 
h^\prime v_d & \gamma (vv_d+ \langle \tilde{N}^0_1 \rangle V)
& 2\gamma v v_u-h^\prime V\\
2\gamma \langle\tilde{N}^0_1\rangle v_u+h^\prime v_d & 
2\gamma (\langle\tilde{N}^0_1\rangle + V^2)+ h^\prime 
\frac{\langle\tilde{N}^0_1\rangle v_d -vV}{v_u} & 
2\gamma v_uV-h^\prime v & 
2\gamma (v\langle\tilde{N}^0_1\rangle + v_dV) \\
\gamma (vv_d+ \langle \tilde{N}^0_1 \rangle V)
& 2\gamma v_u V -h^\prime v & 2\gamma v_u^2 & 
2\gamma v_uv_d + h^\prime \langle\tilde{N}^0_1\rangle \\
2\gamma v v_u-h^\prime V & 
2\gamma (v\langle\tilde{N}^0_1\rangle+v_dV) & 
2\gamma v_dv_u+h^\prime \langle\tilde{N}^0_1\rangle & 
2\gamma (v^2+v_d^2)+h^\prime 
\frac{v_d\langle\tilde{N}^0_1\rangle-vV}{v_u} 
\end{array}\right).\]
\end{widetext}
The analysis shows that only for $h^\prime = 0$ this matrix has two
eigenvalues equal to zero which correspond to the four Goldstone bosons
needed to produce masses for $W^\pm$ and $K^\pm$. So, $h^\prime = 0$ is
mandatory ($h^\prime =0$ is a consequence of the symmetry
$L_{1L}\leftrightarrow L_{2L}$). For the other two eigenvalues one is in 
the TeV scale and the other one at the electroweak mass scale.

\subsubsection{\label{sec:subsub442}The neutral real sector}
For the neutral real sector and in the basis $(\tilde{\nu}_R, 
\tilde{N}_{1R}, \tilde{N}_{2R}, \tilde{N}_{3R}, \tilde{N}_{4R})$ we get 
the following mass matrix:
\begin{equation}
M^2_{real}=\left(\begin{array}{cc} M_{2\times 2} & M_{2\times 3} \\ 
M^T_{2\times 3} & M_{3\times 3}  \end{array} \right),
\end{equation}
where the submatrices are:

\begin{equation}\label{m22}
M_{2\times 2}=
\left(\begin{array}{cc} \gamma v_d^2+4\delta v^2 - 
\frac{h^\prime v_uV}{2v} & 
\gamma v_dV + 4\delta v\langle\tilde{N}_1^0\rangle \\
\gamma v_dV + 4\delta v\langle\tilde{N}_1^0\rangle & 
\gamma V^2+4\delta \langle\tilde{N}_1^0\rangle^2 
- \frac{h^\prime v_uV}{2v}\end{array} 
\right),
\end{equation}
\begin{widetext}
\begin{equation}\label{m23}
M_{2\times 3}= 
\left(\begin{array}{ccc} 
v_dv(4\delta-\gamma) & 2\alpha vV+\gamma v_d \langle\tilde{N}_1^0\rangle
+ h^\prime v_u/2 & 2\beta v v_u+h^\prime V/2 \\
2\alpha\langle\tilde{N}_1^0\rangle v_d+\gamma vV-h^\prime v_u/2 & 
(\gamma-4\delta)vv_d & 2\beta\langle\tilde{N}_1^0\rangle v_u-h^\prime 
v_d/2 \end{array} \right),
\end{equation}

\begin{equation}\label{m33} M_{3\times 3}= \left(\begin{array}{ccc} \gamma
v^2+4\delta v_d^2-\frac{h^\prime vv_u}{2V} & \gamma v
\langle\tilde{N}_1^0\rangle +4\delta v_d V & 2\beta
v_uv_d-h^\prime\langle\tilde{N}_1^0\rangle /2 \\ \gamma v
\langle\tilde{N}_1^0\rangle + 4\delta v_dV &
\gamma\langle\tilde{N}_1^0\rangle^2 +4\delta V^2- \frac{h^\prime vv_u}{2V}
& 2\beta v_uV+h^\prime v/2 \\ 2\beta
v_uv_d-h^\prime\langle\tilde{N}_1^0\rangle /2 & 2\beta v_uV+h^\prime v/2 &
4\eta v_u^2+h^\prime \frac{\langle\tilde{N}_1^0\rangle
v_d-vV}{2v_u}\end{array} \right). \end{equation}
\end{widetext} 
Using the constraints in Eqs. (\ref{minim}), this mass matrix has one
eigenvalue equal to zero which identifies one real Goldstone boson needed
to produce a mass for $K^{0\mu}_I$.  Now, using $h^e\approx 0 ,\;\;  
h^\prime =0$ and with the other values as given before, we get for the
remaining four eigenvalues that two of them are in the TeV scale, other
one is at the electroweak mass scale, while for the lightest CP-even
scalar $h$ we get a tree-level mass smaller than the one obtained in the
MSSM. This result, which is strongly dependent on the value of $h^e$, is
not realistic due to the fact that the radiative corrections have not been
taken into account, but such analysis is not in the scope of the present
work.

\subsubsection{\label{sec:subsub443}The neutral imaginary sector}
For the neutral imaginary sector and in the basis $(\tilde{\nu}_I, 
\tilde{N}_{1I}, \tilde{N}_{2I}, \tilde{N}_{3I}, \tilde{N}_{4I})$ we get 
the following mass matrix:
\begin{equation}
M^2_{imag}=\left(\begin{array}{cc} M^\prime_{2\times 2} & 
M^\prime_{2\times 3} \\ 
M^{\prime T}_{2\times 3} & M^\prime_{3\times 3}  \end{array} \right),
\end{equation}
where the submatrices are
\begin{equation}\label{m22p}
M^\prime_{2\times 2}=
\left(\begin{array}{cc} \gamma v_d^2 - \frac{h^\prime v_uV}{2v} & 
\gamma v_dV \\
\gamma v_dV &  \gamma V^2 - \frac{h^\prime v_uV}{2v}\end{array} 
\right),
\end{equation}

\begin{equation}\label{m23p}
M_{2\times 3}^\prime = 
\left(\begin{array}{ccc} 
\gamma \langle\tilde{N}_1^0\rangle V & -\gamma\langle\tilde{N}_1^0\rangle
v_d-h^\prime \frac{v_u}{2} & -h^\prime \frac{V}{2} \\
-\gamma vV+h^\prime \frac{v_u}{2} & -\gamma\langle\tilde{N}_1^0\rangle V & 
h^\prime \frac{v_d}{2} \end{array} \right),
\end{equation}

\begin{equation}\label{m33p}
M_{3\times 3}^\prime = 
\left(\begin{array}{ccc} 
\gamma v^2 - \frac{h^\prime vv_u}{2V} & 
\gamma v \langle\tilde{N}_1^0\rangle & 
h^\prime\langle\tilde{N}_1^0\rangle /2 \\
\gamma v \langle\tilde{N}_1^0\rangle + & 
\gamma\langle\tilde{N}_1^0\rangle^2 -
\frac{h^\prime vv_u}{2V} & h^\prime v/2 \\
h^\prime\langle\tilde{N}_1^0\rangle /2 & 
-h^\prime v/2 & h^\prime
\frac{\langle\tilde{N}_1^0\rangle v_d-vV}{2v_u}\end{array} \right). 
\end{equation}

Using the constraints in Eqs. (\ref{minim}), this mass matrix has three
eigenvalues equal to zero which identify three real Goldstone bosons (two
of them CP-odd), needed to produce masses for $Z_0^\mu,\; Z^{\prime\mu}_0$
and $K^{0\mu}_R$.

In the limit $h^\prime = 0$, this mass matrix has one eigenvalue in the
TeV scale and four eigenvalues equal to zero that correspond to the three
Goldstone bosons identified for the case $h^\prime\neq 0$, plus an extra
CP-odd scalar of zero mass at tree level.

\subsection{\label{sec:sub45}Masses for Charginos}
The charginos in the model are linear combinations of the
charged leptons and charged gauginos. In the gauge eigenstate basis 
$\psi^\pm = (e^+_1, E^+_1, \tilde{W}^+, \tilde{K}^+, 
e^-_1, E^-_1, \tilde{W}^-, \tilde{K}^-)$ the chargino mass terms in the 
Lagrangian are of the form $(\psi^\pm)^TM\psi^\pm$, where 
\[M=\left(\begin{array}{cc}
0 & M_C^T \\
M_C & 0 \end{array}\right),\]
and 
\begin{equation}
M_C = \left(\begin{array}{cccc}
h^ev_d & -h^ev & 0 & -g_2v_u \\
-h^eV  & h^e \langle \tilde{N}^0_1 \rangle & -g_2v_u & 0 \\
-g_2v & -g_2v_d & M_2  & 0 \\
-g_2\langle \tilde{N}^0_1 \rangle & -g_2V & 0 & M_2 
\end{array}\right).
\end{equation}\\
In the limit $h^e=0$ and $M_2$ very large, this mass matrix is a see-saw
type matrix. The numerical evaluation using the parameters as stated
before produces a tree-level mass for the $\tau$ lepton of the order of 
$1$ GeV, with all the other masses above $90$ GeV.

\section{\label{sec:sec5}General Remarks and Conclusions}
We have built the complete supersymmetric version of the 3-3-1 model in
Ref. \cite{spm} which, like the MSSM, has two Higgs doublets at the
electroweak energy scale (the isospin doublets in $\tilde{L}_{1L}$ and
$\tilde{L}_{3L}$). Since the MSSM is not an effective theory of the model
constructed, exploring the Higgs sector at the electroweak energy scale it
is important to realize that, the MSSM is not the only possibility for two
low energy Higgs doublets.

For the model presented here the slepton multiplets play the role of the
Higgs scalars and no Higgsinos are required, which implies a reduction of
the number of free parameters compared to other models in the literature
\cite{susy331}.

The absence of bilinear terms in the bare superpotential avoids the
presence of possible unwanted $\mu$ terms; in this way the so called $\mu$
problem is absent in the construction developed in this paper.

The sneutrino, selectron and other six sleptons do not acquire masses in
the context of the model, and they play the role of the Goldstone bosons
needed to produce masses for the gauge fields. The right number of
Goldstone bosons is obtained by demanding $h^\prime = m_{L_1L_2}=0$ in
$V_{\hbox{soft}}$.

$h^\prime =0$ in $V_{\hbox{soft}}$ has as a consequence the existence of a
zero mass CP-odd Higgs scalar at tree level. Once radiative corrections
are taken into account we expect it acquires a mass of a few (several?)
GeV, which in any case is not troublesome because, as discussed in Ref.
\cite{nir}, a light CP-odd Higgs scalar not only is very difficult to be
detected experimentally, but also it has been found that in the two Higgs
doublet model type II and when a two-loop calculation is used, a very
light ($\sim 10$ GeV) CP-odd scalar $A_0$ can still be compatible with
precision data such as the $\rho$ parameter, $BR(b \rightarrow s\gamma)$,
$R_b$, $A_b$, and $BR(\Upsilon \rightarrow A_0\gamma)$\cite{larios}.

$h^e=0$ or very small is a necessary condition in order to have a
consistent model, in the sense that it must include a very light neutrino,
with masses for the other spin 1/2 neutral particles larger than half the
$Z^0$ mass. There is not problem with this constraint, because due to the
existence of heavy leptons in the model, $h^e$ is not the only parameter
controlling the charged lepton masses.

We have also analyzed the mass value at tree-level for $h$, the lightest
CP-even Higgs scalar in this model, which is smaller than the lower bound
of the lightest CP-even Higgs scalar in the MSSM, although strongly
dependent on the radiative corrections. This fact is not in conflict with
experimental results due to the point that the coupling $hZZ$ and $hA_0Z$
are suppressed because of the mixing of the $SU(2)_L$ doublet sleptons
with the singlets $\tilde{N}_1^0$ and $\tilde{N}_3^0$.

The recent experimental results announced by the Muon $(g-2)$
collaboration\cite{muong-2} show a small discrepancy between the SM
prediction and the measured value of the muon anomalous spin precession
frequency, which only under special circumstances may be identified with
the muon's anomalous magnetic moment $a_\mu$\cite{feng}, a quantity
related to loop corrections.

Immediately following the experimental results a number of papers appeared
analyzing the reported value, in terms of various forms of new physics,
starting with the simplest extension of the SM to two Higgs
doublets\cite{larios}, or by using supersymmetric extensions, technicolor
models, leptoquarks, exotic fermions, extra gauge bosons, extra
dimensions, etc., in some cases extending the analysis even at two loops
(for a complete bibliography see the various references in \cite{dedes}).
More challenging, although not in complete agreement between the
different authors, are the analyses presented in Refs.\cite{wells} and
\cite{kane} where it is shown how the MSSM parameter space gets
constrained by the experimental results.

Our model, even though different from the MSSM shares with it the property
that very heavy superpartners decouple from the $a_\mu$ value yielding a
negligible contribution. Nevertheless, the model in this paper includes
many interesting new features that may be used for explaining the measured
value of the muon's anomalous precession frequency, as for example a light
CP-odd and a light CP-even scalars which get very small masses at tree
level, but that the loop radiative corrections may rise these masses up to
values ranging from a few GeV to the electroweak mass scale. But an
analysis similar to the one presented in Refs.\cite{wells} and \cite{kane}
is outside the scope of the present study, because in our case it depends
crucially on the predicted values of the Higgs scalar masses, an obscure
matter in supersymmetry. (For example, $a_\mu^{exp}$ can be understood in
the context of our model if the CP-odd scalar has a mass of the order of a
few GeV\cite{larios}, with all the other scalars and supersymmetric
particles acquiring masses larger than the electroweak mass scale.
Similarly, the light CP-even Higgs boson $h$ with enough suppressed $hZZ$
and $hA_0Z$ couplings can contribute significantly to
$a_\mu$\cite{dedes}).

The idea of using sleptons as Goldstone bosons is not new in the
literature \cite{kaku}, but as far as we know there are just a few papers
where this idea is developed in the context of specific models, all of
them related to one family structures \cite{rizo}.

The model can be extended to three families, but the price is high since
nine $SU(3)_L$ triplets of leptons with their corresponding sleptons are
needed, which implies the presence of nine $SU(2)_L$ doublets of Higgs
scalars. An alternative is to work with the three family structures
presented in Refs.\cite{fp, long}.

In conclusion, the present model has a rich phenomenology and it deserves
to be studied in more detail.

\section*{ACKNOWLEDGMENTS} 
Work partially supported by Colciencias in Colombia and by CODI in the U.  
de Antioquia. L.A.S. acknowledges partial financial support from U.  de
Antioquia. We thank E. Nardi and M. Losada for critical readings of the
original manuscript.

\end{document}